\documentclass[twocolumn]{aastex63}

\usepackage{graphicx}% Include figure files
\usepackage{dcolumn}% Align table columns on decimal point
\usepackage{bm}% bold math
\usepackage{acronym}
\usepackage{amsmath}
\usepackage[cal=boondox]{mathalfa}
\usepackage{aas_macros}
\acrodef{GW}{gravitational wave}
\acrodef{PE}{Parameter Estimation}
\acrodef{MCMC}{Markov Chain Monte Carlo}
\acrodef{DOF}{degrees of freedom}
\acrodef{AGN}{Active Galactic Nuclei}

\newcommand{\D}{\mathrm{d}}

\newcommand{\checkme}[1]{{#1}}
\newcommand{\ratem}{\checkme{25.8}}
\newcommand{\rateu}{\checkme{11.5}}
\newcommand{\ratel}{\checkme{8.8}}

\graphicspath{{figures/}}

\begin{document}

\preprint{APS/123-QED}

\title{Applying Cosmological Principle to Better Probe the Redshift Evolution of Binary Black Hole Merger Rate}

\correspondingauthor{Vaibhav Tiwari}
\email{tiwariv@cardiff.ac.uk}

\author[0000-0002-1602-4176]{Vaibhav Tiwari}
\affiliation{Gravity Exploration Institute \\
School of Physics and Astronomy, \\ 
Cardiff University, Queens Buildings, The Parade \\ 
Cardiff CF24 3AA, UK.}

\date{\today}% It is always \today, today,

\begin{abstract}
Gravitational waves inform about the probable distances at which an observed signal originated. This information when combined over multiple observations is used in the modeling of the redshift evolution of the merger rate. This is an important aspect of binary black hole population analysis which is expected to have close ties with the star formation history of the universe or dynamical evolution of star clusters. At the least, it can probe the time delay between the star formation and merger of remnants. However, due to the degeneracy between the inclination angle of the binary and the luminosity distance, the measured luminosity distance has large uncertainty that limits our ability to precisely measure the redshift distribution of mergers. In this letter, we show that by imposing the cosmological principle it is possible to suppress this uncertainty and better discriminate between different distributions modeling the redshift evolution of binary black hole merger rate. Additionally, we show that by making a comparison with an analysis that does not make such imposition we can probe the cosmological principle.
\end{abstract}


\section{Introduction} \label{sec:intro}
A \ac{GW} signal contains enough information to determine the luminosity distance to the binary \citep{1986Natur.323..310S}. Distance distribution of multiple observations can be combined to infer the redshift evolution of the merger rate \citep{2018ApJ...863L..41F, 2019ApJ...882L..24A, 2020arXiv200807014R, o3a_rnp}\footnote{We fix the cosmological parameters to their Planck 2015 values \citep{2016A&A...594A..13P}.}. This is an important aspect in the population analysis of merging black holes as it has close ties with the star formation history of the universe or the dynamical evolution of star clusters. At the least, it can probe the time delay between the star formation and merger of remnants. For isolated binaries the time delay between formation of a stellar mass binary and merger of the remnants is due to the evolution of components in a binary, the common envelope phase, and due to orbital evolution caused by the emission of the \ac{GW}s \citep{2014LRR....17....3P, 2015ApJ...806..263D}. Binaries are also formed in star cluster after components segregate to the core and form and tighten binaries due to many body dynamics \citep{2000ApJ...528L..17P}. The usual effect of delay results in a shallower evolution of the merger rate of remnants with the redshift compared to the redshift evolution of the star formation rate \citep{2020ApJ...898..152S}.

In this letter we reconstruct the evolution of the merger rate with redshift using observations in GWTC-2 occurring during the first two observing runs and the first half of the third run (O1, O2, and O3a) \citep{losc}. We extend the flexible mixture-model framework VAMANA\citep{vamana} to model the evolution as a power-law function independent of other signal parameters. Our work differs from previous analysis in using a flexible Gaussian mixture model that admits chirp mass as a signal parameter.  Additionally and more importantly we show that the uncertainty on reconstruction can be reduced by imposing homogeneity and isotropy of the universe. Moreover, by comparing with an independent analysis that does not make such imposition we probe the cosmological principle. We describe the method and analysis in section \ref{sec:method} and present results in section \ref{sec:results}.
\section{Method and Analysis}
\label{sec:method}
The methodology to model population properties of merging compact binaries has been discussed in multiple publications \citep{2019MNRAS.486.1086M, 2018ApJ...868..140T, 2019PASA...36...10T}. Following \citep{2018ApJ...868..140T}, the posterior on model hyper-parameters is given by equation \ref{eq:pop_bayes},
\begin{eqnarray}
p(\lambda | \{\bm{d}\}) &\propto&
\prod_{i=1}^{N_\mathrm{obs}} \frac{\int \D\bm{\theta}\;p(\bm{d}_{i} | \bm{\theta} )\;p (\bm{\theta} | \lambda)}{\int \D\bm{\theta}\;p_{\mathrm{det}}(\bm{\theta})\; p(\bm{\theta}|\lambda)}\,p(\lambda) \nonumber \\
&\equiv& e^{\mathcal{L}}\,p(\lambda)
\label{eq:pop_bayes}
\end{eqnarray}
where $\bm{d} \equiv \{ \bm{d_0}, \cdots, \bm{d}_{N_\mathrm{obs}}\}$ is the set of observations, $\lambda$ is the population model, $\bm{\theta}$ are the signal parameters admitted in the population analysis and $p_\mathrm{det}(\bm{\theta})$ encodes the probability of an event with signal parameters $\bm{\theta}$ to be observed with confidence. $p(\lambda)$ is the prior probability on the model hyper-parameters and $\mathcal{L}$ is the log-likelihood. The analysis samples the posterior $p(\lambda | \{\bm{d}\})$ using a \ac{MCMC} method and thus does not require the normalisation constant for equation \ref{eq:pop_bayes}.
In practice equation \ref{eq:pop_bayes} is estimated using discrete samples. \ac{PE} analysis samples $p(\bm{d}_{i} | \bm{\theta} )$ for a population model $p (\bm{\theta} | \lambda_{\mathrm{PE}})$ \citep{2015PhRvD..91d2003V}, and large scale injection campaign are performed to estimate the sensitivity of the detector network for a population model $p(\bm{\theta} | \lambda_{\mathrm{inj}})$. Both the numerator and the denominator are then calculated for a target population $p (\bm{\theta} | \lambda)$ using importance sampling \citep{2018CQGra..35n5009T, 2019MNRAS.486.1086M}.

Probability distribution in equation \ref{eq:pop_bayes} is extended to include the merger rate by incorporating the Poisson term \citep{extended_lkl}
\begin{eqnarray}
p(N_\mathrm{obs} | \lambda, \{\bm{d}\}) = \frac{\mu^{-N_\mathrm{obs}}\mathrm{e}^{-\mu}}{N_\mathrm{obs}!} \\
\mu = \int R(z)\,p(\bm{\theta}|\lambda)\frac{1}{1+z}\frac{\mathrm{d}V_c}{dz}\, p_{\mathrm{det}}(\bm{\theta})\; \mathrm{d}\bm{\theta},
\label{eq:poisson}
\end{eqnarray}
where $\D V_c/\D z$ is the differential co-moving volume and $\mu$ is the expected number of observations at population averaged merger rate $R(z) $ for the population model $p(\bm{\theta}|\lambda)$ at redhsift $z$. Often prior on merger rate is chosen to be uniform-in-log that is scale invariant. On marginaling over the merger rate the posteriors of other hyper-parameters remain unaffected and equivalent to am \emph{unextended} analysis \citep{uniforminlog}. Thus Poisson term carries no information.

The redshift evolution of the merger rate can be probed by including a phenomenological model for the redshift evolution of the merger rate. For example, some works have used a phenomenological distribution that is also used in modeling the star formation history \citep{2014ARA&A..52..415M}
\begin{equation}
    R(z) = \frac{R\;(1 + z) ^ \kappa}{1 + [(1 + z)/(1 + z_p)]^\beta},
    \label{eq:pz}
\end{equation}
where $R$ is the population averaged merger rate at $z = 0$ and $z_p$ is the redshift at which the peak star-formation occurs. The ground based gravitational wave detectors have not yet reached sensitivity to observe mergers occurring near peak star-formation \citep{2020LRR....23....3A} therefore current analysis only measure the leading order term which is the powerlaw with exponent $\kappa$ in the numerator of equation \ref{eq:pz} \citep{2018ApJ...863L..41F, 2019ApJ...882L..24A, 2020ApJ...896L..32C, 2020arXiv200807014R, o3a_rnp}. 

In a previous article, we introduced VAMANA a flexible population analysis framework \citep{vamana}. VAMANA models the chirp mass, mass-ratio, and aligned spin distribution using a mixture model. The analysis explores all distributions – expressible as the sum of weighted  Gaussians – that are within a  distance measure from the reference chirp mass distribution. The reference chirp mass distribution is a simple power-law that obtains the maximum value of $\mathcal{L}$. We refer the interested reader to section IIIB of \cite{vamana} for a detailed description of the model. Equation \ref{eq:pz} is independent of the other signal parameters and we extend VAMANA to model redshift evolution of merger rate by multiplying the existing population model by the density
\begin{equation}
p(z\,|\,k) \propto \frac{1}{1+z}\frac{\mathrm{d}V_c}{dz} (1 + z)^\kappa.
\label{eq:pz1}
\end{equation}
The normalisation constant in the previous equation can be ignored as it cancels from the numerator and the denominator of equation \ref{eq:pop_bayes}. Having incorporated this model, we also maximise on $\kappa$ when obtaining the reference population.

The phenomenological model in equation \ref{eq:pz1} has only one hyper-parameter, in-spite of that, the measured credible intervals on $\kappa$ are large \citep{o3a_rnp}, and thus there is a large uncertainty on how steep or how shallow is the evolution of the merger rate with the redshift. The primary source of the uncertainty in $\kappa$ is due to the uncertainty in the measurement of luminosity distance, which occurs because of degeneracy between the luminosity distance and the inclination angle of a binary \citep{2019ApJ...877...82U}. Loosely speaking, there isn't enough information in the analysis to precisely model the redshift evolution of the merger rate. However, this information can be collected over a large number of observations and a precise inference can be made.

It is also possible to inject information in the analysis. In particular, no where in the population analysis we demand the universe to be isotropic and homogeneous. In fact, we only supply this as a prior information when estimating parameters from the observed signals \citep{2015PhRvD..91d2003V}. Cosmological principle is a notion that universe is isotropic and homogeneous \citep{1972gcpa.book.....W, 1993ppc..book.....P} and is well supported by the data \citep{2005pfc..book.....M}. One way to enforce it is by binning the redshift and extending the likelihood in equation \ref{eq:pop_bayes} by the probability
\begin{equation}
    p(\{N^1_\mathrm{obs}, N^2_\mathrm{obs}, \cdots, N^b_\mathrm{obs}\} | \lambda, \{\bm{d}\}) = \frac{\mu^{-N_\mathrm{obs}}\mathrm{e}^{-\mu}}{\prod_{j = 1}^{j = b} N^j_\mathrm{obs}!},
    \label{eq:bpoisson}
\end{equation}
where $b$ is the number of bins and $ N_\mathrm{obs} = \sum_{j = 1}^{j = b} N^j_\mathrm{obs}$. Thus, in addition to the total number of observation as predicted by the population model and population averaged merger rate, we are also interested in their distribution over the bins. The numerator in equation \ref{eq:poisson} and \ref{eq:bpoisson} are the same but denominator in equation \ref{eq:bpoisson} acts as a goodness of fit with an isotropic and homogeneous universe. We are not applying cosmological principle in the strictest sense as our bin sizes are of the order of few hundred mega-parsecs and it is conceivable to have an anisotropic universe that can reproduces the bin counts, $N^j_\mathrm{obs}$, as predicted by an isotropic universe. To impose isotropy strictly one has to create bins in the sky \citep{2020arXiv200302919S, 2020arXiv200611957P}.

To count the number of observation in a bin we introduce the flag $g_{ij}$ which assigns the $i^{\mathrm{th}}$ observation  to the $j^{\mathrm{th}}$ bin,
\begin{equation}
g_{ij}=
\begin{cases}
1\quad\mathrm{for} \quad i = j,\\
0 \quad \mathrm{otherwise},
\end{cases}
\end{equation}
and the corresponding function,
\begin{equation}
g_{ij}(z)=
\begin{cases}
1\quad\mathrm{for} \quad z^j_{\mathrm{\min}} < z \leq z^j_{\mathrm{\max}},\\
0 \quad \mathrm{otherwise}.
\end{cases}
\end{equation}
where $z^j_{\mathrm{\min}}$ and $z^j_{\mathrm{\max}}$ are the boundaries of the $j^{\mathrm{th}}$ bin. The number of observation in bins is then
\begin{equation}
    \{N^1_\mathrm{obs}, N^2_\mathrm{obs}, \cdots, N^b_\mathrm{obs}\} = \{\sum_{i=1}^{N_\mathrm{obs}} g_{i1}, \cdots, \sum_{i=1}^{N_\mathrm{obs}} g_{ib}\}.
\end{equation}
Along with the probability in equation \ref{eq:bpoisson} we also need $p(g_{ij}|\lambda, \bm{\bm{d}_i})$ which is the probability that $i^{\mathrm{th}}$ observation originated from the $j^{\mathrm{th}}$ bin. This can be easily calculated by re-weighting the redshift estimates \citep{2019MNRAS.486.1086M},
\begin{equation}
    p(g_{ij}|\lambda, \bm{\bm{d}_i}) = \frac{\sum_k g_{ij}(z_k^i)\,p(\theta_k^i|\lambda) / p (\theta_k^i | \lambda_{\mathrm{PE}})}{\sum_k p(\theta_k^i|\lambda) / p (\theta_k^i | \lambda_{\mathrm{PE}})},
    \label{eq:pgij}
\end{equation}
where $\theta_k^i$ are the sampled parameters for the $i^{\mathrm{th}}$ signal.

We perform two analysis, one where we use one bin and calculate the Poisson probability using equation \ref{eq:poisson} and the other where we create four redshift bins and calculate the Poisson probability using equation \ref{eq:poisson} and \ref{eq:pgij}. We chose [0, 0.19, 0.27, 0.41, 1.50] as the bin edges. With this choice the expected number of observations in the bins are approximately equal for the reference population. We use a uniform prior on $\kappa$ in the range [-6, 6]. For the second analysis, we use a uniform prior on $g_{ij}$. All the \ac{PE} samples and injection campaign's data we have used in this analysis is publicly available \citep{losc}. All our settings, priors, and event selection remain intact and as described in \citep{2020arXiv201104502T}. 
\section{Results}
\begin{figure*}
    \centering
    \includegraphics[width=\textwidth]{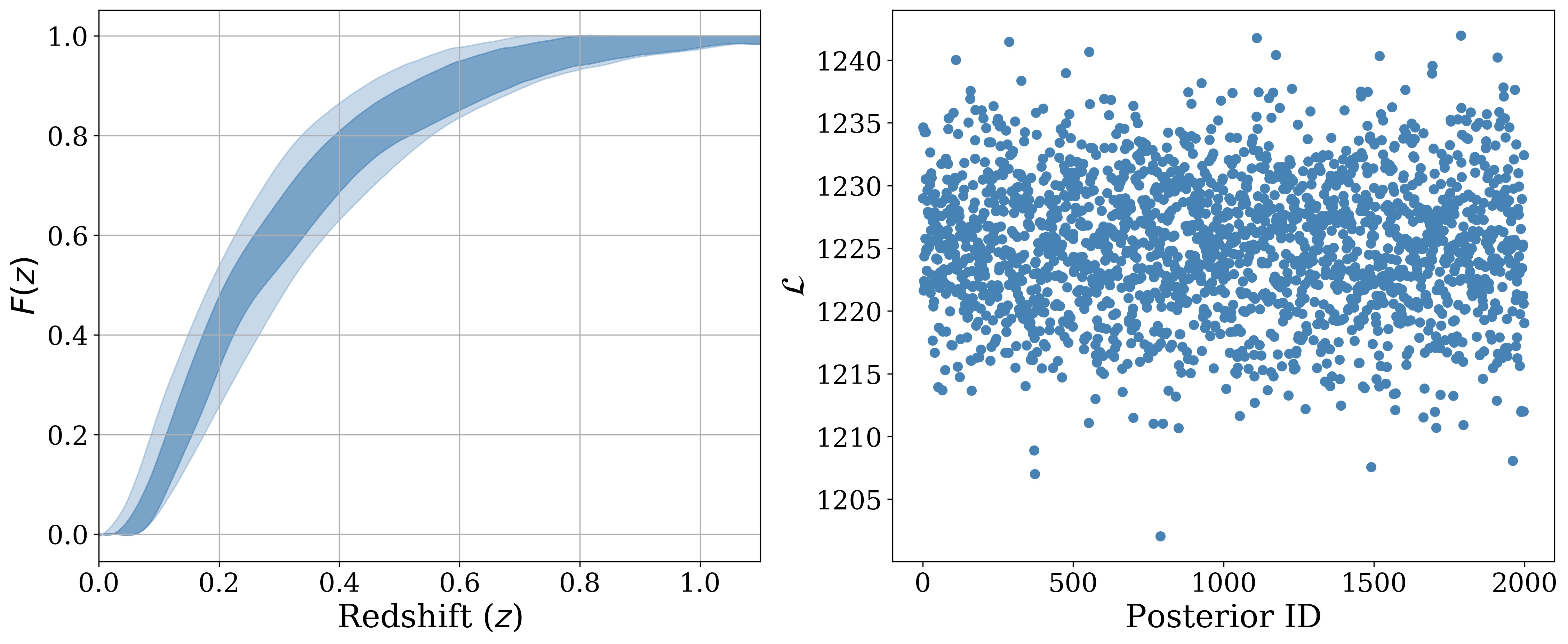}
    \caption{Left) The light blue band is the 90\% confidence of the cumulative probability of the posterior predictive obtained after applying selection effects to the reconstructed redshift evolution of the merger rate. The dark blue band is the 90\% confidence obtained by bootstrapping various realisations of the observed data. Each realisation of the observed data is generated by first performing importance sampling on the redshift estimate of the observations using posterior values of $\kappa$ and then selecting one data point from each observation. The observed data is enclosed within the 90\% confidence of the posterior's prediction, right) The natural log of the likelihood defined in equation 2 showing no visible trend in their values indicates proper convergence of the sampler. Both the plots correspond to the four bin case.}
    \label{fig:sanity}
\end{figure*}

In this section, we compare the results obtained for the two analyses. We perform a few sanity checks to verify the correctness of the analysis. Each posterior in our analysis is equally probable to be the true mass, spin, and the redshift distribution. For a reconstructed population, $p(\bm{\theta}|\lambda)$,  corresponding to each posterior $\lambda$, we apply selection effects and generate multiple realisation of expected observations from this population model. Each prediction is chosen to have the same number of data points as the number of observations used in the analysis. Moreover, we can also generate multiple realisations of the observed data. We perform importance sampling on each observation using the population model corresponding to each posterior and randomly select one sample from each observation. Figure \ref{fig:sanity} plots the cumulative distributions for the predicted and the observed data. The 90\% credible interval of the observed data lies within the 90\% credible interval of the predicted data.
\label{sec:results}

\begin{figure*}
    \centering
    \includegraphics[width=\textwidth]{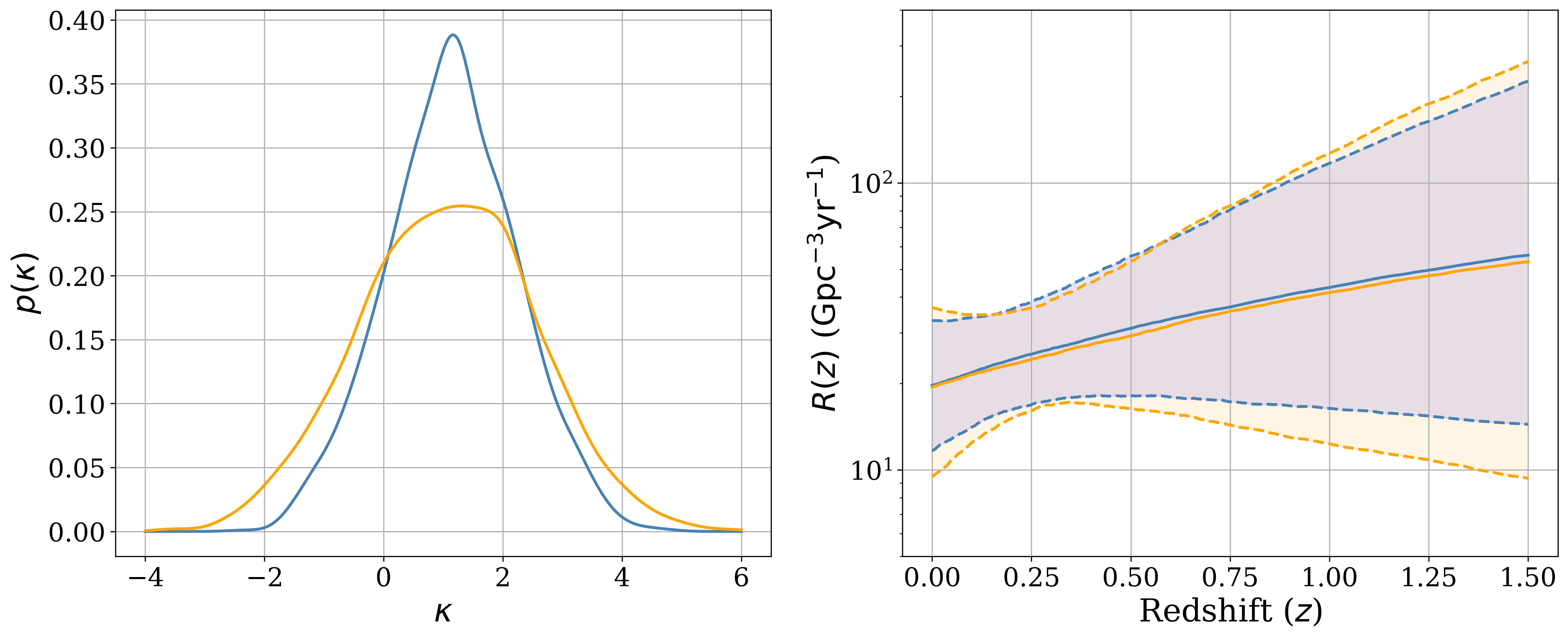}
    \caption{Left) The posterior on $\kappa$. The blue curve is for the four bin case and the orange curve|
    is for the one bins case. The width of the credible interval shrinks from [-1.37, 3.40]  to [-0.68, 2.97] due to redshift binning, right) The redshift evolution of the merger rate. Binning reduces the width of the merger rate interval at redshift $z$ = 1.50 by a factor of 2.}
    \label{fig:zevol}
\end{figure*}

Figure \ref{fig:zevol} plots the posterior on $k$. For the one bin case we measure $k$ to be $1.09^{+2.30}_{-2.46}$ but this changes to  $1.15^{+1.84}_{-1.83}$ for the four bin case. The credible interval reduces by 1.1. The posterior on $\kappa$ is sensitive to the choice of prior on the mass distribution. VAMANA models the chirp mass as a sum of weighted Gaussians and we choose the location of these Gaussians to have a uniform in log prior. However, if we change the prior on the locations of the Gaussians to be uniform, the mean value of $\kappa$ reduces to around 0.5. 

In \cite{2020arXiv201104502T} we reported a merger rate of $\ratem^{+\rateu}_{-\ratel}\,\mathrm{Gpc}^{-3} \mathrm{yr}^{-1}$ for a rate non-evolving in redshift. After allowing a powerlaw evolution for the merger rate we obtain $20.80^{+16.00}_{-11.40} \mathrm{Gpc}^{-3}\mathrm{yr}^{-1}$ for the one bin case and $20.74^{+12.39}_{-9.09} \mathrm{Gpc}^{-3}\mathrm{yr}^{-1}$ for the four bin case. The mean value of $\kappa$ is positive that results in an increase in the sensitive volume ($\kappa = 0$ for non-evolving merger rate) and a reduced merger rate. 
\begin{figure}
    \centering
    \includegraphics[width=0.5\textwidth]{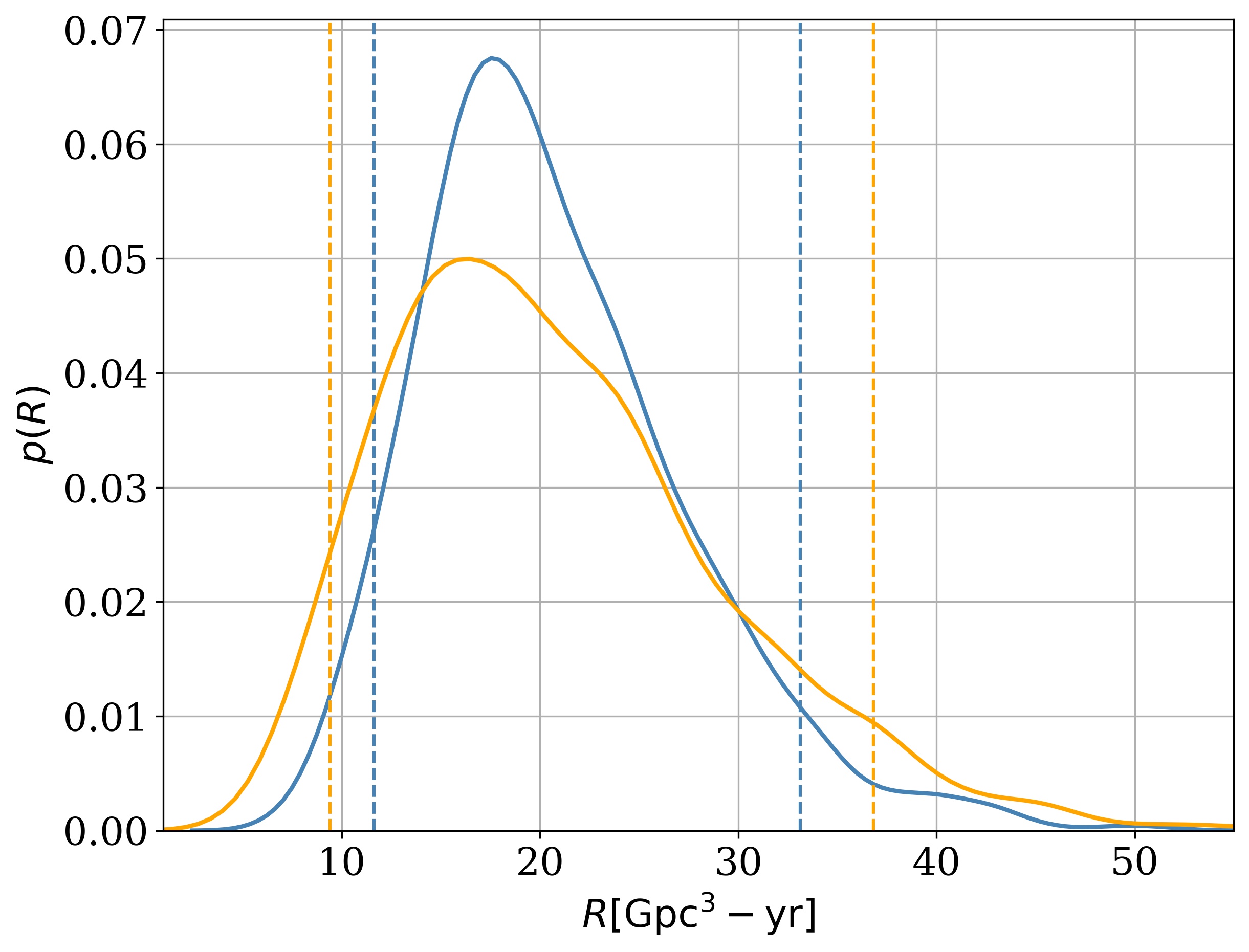}
    \caption{The rate posterior at redshift zero. The orange curve is posterior for the one bin case and the blue curve for the four bins case. Binning reduces the credible interval from [9.4, 36.8] to [11.6, 33.1] $\mathrm{Gpc}^{-3}\mathrm{yr}^{-1}$.
    }
    \label{fig:rate}
\end{figure}
\subsection{Cosmological Principle}
\begin{figure*}
    \centering
    \includegraphics[width=\textwidth]{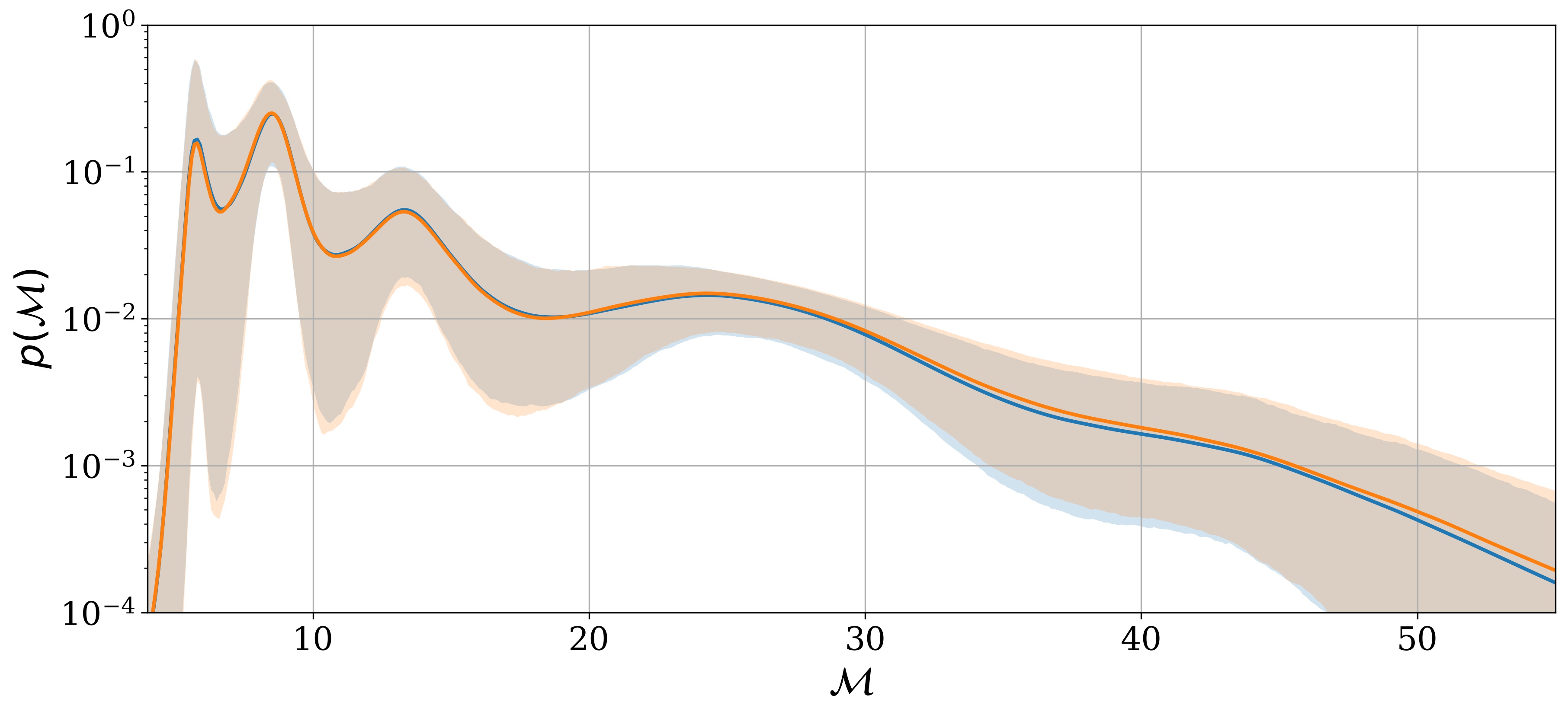}
    \caption{The reconstructed chirp mass distribution for the two analysis and the 90\% credible interval. Orange plot is the reconstruction for the one bin analysis and blue plot is the reconstruction for the four bin analysis.}
    \label{fig:compare_mchirp}
\end{figure*}

The two analysis we perform contrast in the imposition of the cosmological principle. We should stress again that we are not applying the cosmological principle in the strictest sense. In fact our analysis is only sensitive to the relative distribution of observations in the redshift bins which themselves have width of many hundreds of mega-parsecs. However, we can still probe any substantial departure from the homogeneous and isotropic distribution of the signals. In the event cosmological principle is violated, the four binned analysis will not explain the data as well as the one binned analysis. The usual way to perform model selection is by comparing the evidences from the two analysis. But as we use Metropolis-Hastings sampling \citep{10.1093/biomet/57.1.97} we can not calculate the evidence. However, as probably many hundreds of events are required to properly probe the cosmological principle, in this asymptotic limit we expect the mean of the reconstructed distribution to be very close to the true distribution \citep{judith2011}. We thus use a Bayes factor calculated using the mean of the reconstructed distribution.
\begin{eqnarray}\label{eq:odds_ratio}
\mathfrak{R} \equiv \frac{p(\{\bm{d}\} | f_1)}{p(\{\bm{d}\} | f_2 )} = 
\left[\frac{V(f_1)}{V(f_2)}\right]^{-N} \nonumber\\\prod_{i=1}^{N} 
\left[\frac{ \int \D \bm{\theta} p(\bm{d}_i | \bm{\theta})\;p(\bm{\theta}|f_1) }{
\int \D \bm{\theta} p(\bm{d}_i | \bm{\theta})\;p(\bm{\theta}|f_2)}\right],
\label{eq:bf}
\end{eqnarray}
where the probability distribution $p(\bm{\theta}|f)$ is the integral of the population model on the hyper-parameter posterior,
\begin{equation}
    f(\bm{\theta}) = \frac{1}{n}\sum_k^n p(\bm{\theta} | \lambda_k),
\end{equation}
and $n$ in the number of posterior samples. Here we are giving equal prior probability to the distributions $f_1$ and $f_2$. Using 39 observations we calculate $\mathfrak{R}$ to be 1.02. In figure \ref{fig:compare_mchirp} we show the reconstructed chirp mass distribution for the two analysis. The reconstructed mean are nearly identical. The very small difference is most probably due to sampling error and due to the use of different waveform models used in correcting for the selection effects and calculation of Poisson term, and in estimating the signal parameters.

\section{Conclusion}
\label{sec:conclusion}
In the population analysis of merging binary black holes cosmological principle is not explicitly imposed. In this article, we showed that by creating multiple bins in redshift and counting the number of observations in each bin we can loosely demand the isotropy and homogeneity of the universe. We extend VAMANA the flexible mixture model framework to include redshift as a population parameter. We apply our analysis on the publicly available data and show that by imposing the cosmological principle we can substantially improve the inference on the redshift evolution of merger rate. We can better discriminate between different distributions modeling the redshift evolution of the black hole merger rate. In particular, we plan to apply our increased discrimination in probing the mass dependence of the redshift evolution of merger rate or evolution of the mass distribution with the redshift. Additionally, we suggest that we can probe the cosmological principle by comparing an analysis that imposes the isotropy and homogeneity of the universe to an analysis that does not impose the isotropy and homogeneity of space.

\section*{Acknowledgement}
This work was supported by the STFC grant ST/L000962/1.

We are grateful for the computational resources provided by  Cardiff  University and funded by an STFC grant supporting UK Involvement in the Operation of Advanced LIGO. We are also grateful for computational resources provided by the Leonard E Parker Center for Gravitation, Cosmology, and Astrophysics at the University of Wisconsin-Milwaukee and supported by National Science Foundation Grants PHY-1626190 and PHY-1700765

This research has made use of data, software, and/or web tools obtained from the Gravitational Wave Open Science Center (https://www.gw-openscience.org/), a service of LIGO Laboratory, the LIGO Scientific Collaboration, and the Virgo Collaboration. LIGO Laboratory and Advanced LIGO are funded by the United States National Science Foundation (NSF) as well as the Science and Technology Facilities Council (STFC) of the United Kingdom, the Max-Planck-Society (MPS), and the State of Niedersachsen/Germany for support of the construction of Advanced LIGO and construction and operation of the GEO600 detector. Additional support for Advanced LIGO was provided by the Australian Research Council. Virgo is funded, through the European Gravitational Observatory (EGO), by the French Centre National de Recherche Scientifique (CNRS), the Italian Istituto Nazionale della Fisica Nucleare (INFN) and the Dutch Nikhef, with contributions by institutions from Belgium, Germany, Greece, Hungary, Ireland, Japan, Monaco, Poland, Portugal, Spain.

%\clearpage
\bibliography{references}% Produces the bibliography via BibTeX.s

\end{document}